# Comments on "Analysis of permanent magnets as elasmobranch bycatch reduction devices in hook-and-line and longline trials"

**Joshua M. Courtney and Michael W. Courtney**[*]

BTG Research, P.O. Box 62541, Colorado Springs, Colorado, United States of America, 80962
*E-mail: Michael_Courtney@alum.mit.edu

**Abstract**
A recent study (Fish. Bull. 109:394–401 (2011)) purportedly tests two hypotheses: 1. that the capture of elasmobranchs would be reduced with hooks containing magnets in comparison with control hooks in hook-and-line and longline studies. 2. that the presence of permanent magnets on hooks would not alter teleost capture because teleosts lack the ampullary organ. Review of this paper shows some inconsistencies in the data supporting the first hypothesis and insufficient data and poor experimental design to adequately test the second hypothesis. Further, since several orders of teleosts are known to possess ampullary organs and demonstrate electroreception, grouping all teleosts in a study design or data analysis of magnetic hook catch rates is not warranted. Adequate tests of the hypothesis that permanent magnets or magnetized hooks do not alter teleost capture requires a more careful study design and much larger sample sizes than O'Connell et al. (Fish. Bull. 109:394–401 (2011)).

A recent study (O'Connell et al., 2011) purportedly tests two hypotheses. The first hypothesis is that "the capture of elasmobranchs would be reduced with hooks containing magnets in comparison with control hooks in hook-and-line and longline studies." Both the study design and the data presented seem adequate to support this hypothesis for the species where support is claimed (*Rhizoprionodon terraenovae*, *Mustelus canis*, *Carcharhinus limbatus,* and *Dasyatis americana*), but there are some inconsistencies in the data and analyses that are presented. First, Table 3 lists the total number of elasmobranchs caught on all treatments in the hook-and-line study as 147, but then lists 119 caught on the control and 57 caught on the magnet treatment, a total of 176. Summing the elasmobranchs in the *n* column suggests a total of 300 elasmobranchs on all treatments, which would imply a total of 124 elasmobranchs caught on the procedural control (sham, lead weight treatment).

    The second and more significant issue with the analyses in support of the first hypothesis is the P value used to determine significance in the study. Both Table 2 and Table 3 state, "Asterisks indicate significant ($P<0.005$) differences between control and magnetic treatments in chi-square analysis." However, the lines marked with asterisks in Table 2 have P values of $P=0.0396$ and $P=0.0348$ showing that these groups were either not significant, or that significance was determined with a P threshold different from that stated in Table 2. Similarly, $P=0.0067$ is reported for *Mustelus canis* in Table 3. While this would be considered significant with some common P thresholds for determining significance, it is not consistent with the stated threshold $P<0.005$.

    The most objectionable issue with the paper is that neither the study design nor the quantity of data is adequate for testing the second hypothesis that "that the presence of permanent magnets on hooks would not alter teleost capture because teleosts lack the ampullary organ." In spite of the fact that neither the discussion nor the conclusion contain any consideration of the second hypothesis or the results in teleost species, the abstract makes the unsupported assertion that "Teleosts, such as red drum (*Sciaenops ocellatus*), Atlantic croaker (*Micropogonias undulatus*), oyster toadfish (*Opsanus tau*), black sea bass (*Centropristis striata*),

and the bluefish (*Pomatomas saltatrix*)[sic], showed no hook preference in either hook-and-line or longline studies." Only four teleosts were caught in the longline studies, apparently all red drum. The 2-2 split between the magnet and control groups is insufficient to conclude that red drum show no hook preference. A sub-sample of only 4 fish is recognized as "data deficient" when discussing *Carcharhinus limbatus*; however, somehow significance is attributed to the teleost sample size of only 4 red drum in the longline study. Furthermore, the sub-sample of 11 teleosts, divided among four species in the hook-and-line study is inadequate to test the hypothesis predicting no difference in teleost capture rates between the magnet and the control treatments. Another way to look at the issue is to consider whether the aggregated teleosts in the hook-and-line study show a different level of avoidance for the magnet than the elasmobranchs, which (in aggregate) were caught on the control 119 times and were caught on the magnetic hook 57 times. The 6 to 5 control to magnet outcome of the aggregated teleosts fails to reject the hypothesis of "no difference" from the elasmobranchs with $\chi^2 = 0.798$.

In addition to the statistical issues regarding valid tests of the hypothesis, there is the more fundamental biological issue that it is long established that numerous species of teleosts, including several orders, possess electroreceptive capabilities (Kramer, 1996). For example, sea catfish are known to possess the ability to detect electromagnetic fields (Bretschneider and Peters, 1992). However, it is not known what level of fields can be detected by sea catfish; nor is it known if the presence of electromagnetic fields at detectable levels would act as an attractant or a deterrent to a bait presented on a hook. The principal of electromagnetic induction ensures that a magnet placed in salt water will create an induced electric field if the water is moving relative to the magnet. Consequently, the design of the longline study only testing during slack tides is lacking. Similarly, a valid hook-and-line study should analyze the data separately for slack tides and for moving water.

The findings of Bullock and Northcutt (1982) opened up the possibility that "electroreception might turn up anywhere among hundreds of fish families, especially among teleosts … it will not necessarily be homologous to previous known examples" (Bullock 1999). Consequently, the scientific method demands that electroreception and insensitivity to the presence of permanent magnets be demonstrated at a level of taxa below the ordinal and not inferred for broad classes of teleosts based on theoretical considerations or limited data since "most of the 30 orders of fishes not known to have electroreception have probably not been adequately examined … the task is much larger than sampling 30 orders" (Bullock and Northcutt 1982). Grouping of all teleosts is unwarranted. The hypothesis of teleost insensitivity to magnetic hooks should be tested with adequate sample sizes for a number of taxa under varying conditions including moving water. Furthermore, the statistical test(s) applied should be valid for rejecting the hypothesis of a difference rather than merely failing to reject the null hypothesis. Alternatively, one can design the experiment to attempt to reject the hypothesis that the catch rate of a teleost group or species is no different than a given group of elasmobranchs.

**Note**
This manuscript was originally submitted to Fishery Bulletin on 15 June 2013, but the editor informed us that comments on published articles are not considered for publication, even if they bring to light obvious errors in the published manuscript. The editor of Fishery Bulletin subsequently solicited an errata from the paper's authors which was subsequently published, but no acknowledgment was made to us for bringing the obvious errors to their attention. Surprisingly, in their errata, the authors of the original paper still assert that all the paper's conclusions are correct. Some of the authors of the original paper have commercial interests in magnetic hooks and other shark "deterrents." The authors of this comment have no commercial interests in fishing products and no support or pending support from manufacturers of such products. We have recently completed a study showing that magnetic hooks significantly effect catch rates in sea catfish, *Ariposis felis*. The paper reporting these results is currently under review.

## References

Bretschneider F., and R.C. Peters.
    1992. Transduction and transmission in ampullary electroreceptors of catfish. Compar. Biochem. Physiol. 103: 245-252.

Bullock T.H., Northcutt RG.
    1982. A New Electroreceptive Teleost: *Xenomystus nigri* (Osteoglossiformes Notopteridae). J. Compar. Physiol. 148:345-352.

Bullock T.H.
    1999. The Future of Electroreception and Electrocommunication. J. Exp. Biol. 202:1455-1458.

Kramer B.
    1996. Electroreception and Communication in Fishes. Ulm: G Fischer. 119 p.

O'Connell C.P., D.C. Abel, E.M. Stroud, and P.H. Rice.
    2011. Analysis of permanent magnets as elasmobranch bycatch reduction devices in hook-and-line and longline trials. Fish. Bull. 109: 394-401.